\documentclass[a4paper,english,reprint, aps,onecolumn,twocolumn]{revtex4-2}
\usepackage{amsmath,amssymb,graphicx} 
\usepackage[LGR,T1]{fontenc}
\usepackage[latin9]{inputenc}
\usepackage{textcomp}
\usepackage{esint}

\usepackage{hyperref}
\usepackage{color}
\usepackage{booktabs}

\makeatletter

\begin{document}
\title{Dynamic Synchronization and Resonance as a Universal Origin of 1/f Fluctuations \\
- Amplitude Modulation Across Music and Nature
}
\author{Akika Nakamichi}
\email{nakamichi@cc.kyoto-su.ac.jp}
\affiliation{General Education, Kyoto Sangyo University ~\\
Motoyama Kamigamo Kita-ku, Kyoto 603-8555 Japan }

\author{Izumi Uesaka}
\affiliation{Department of Physics, Kyoto Sangyo University ~\\
Motoyama Kamigamo Kita-ku, Kyoto 603-8555 Japan }

\author{Masahiro Morikawa}
\email{hiro@phys.ocha.ac.jp}
\affiliation{RIKEN, Wako-Saitama 351-0198 Japan }
\date{\today}
\begin{abstract}
We propose a universal physical mechanism for the emergence of 1/f fluctuations, observed across a wide range of systems. In particular, we verify this on acoustic cases. The mechanism is based on amplitude modulation (AM) and demodulation (DM), where the 1/f spectral law arises not in the raw waveform but in its demodulated amplitude envelope.
Two distinct yet complementary processes generate the required AM: (i) stochastic synchronization among oscillators, modeled via an extended Kuramoto framework that captures perpetual synchronization-desynchronization cycles, and (ii) frequency-selective resonance, modeled by spectral accumulation of eigenmodes in acoustic or structural environments. Numerical simulations demonstrate that both mechanisms, acting separately or in combination, robustly produce 1/f spectra over several decades when DM is applied, and that the classical Kuramoto critical point is not necessary for their emergence.
We demonstrate the cross-domain relevance of this AM/DM framework through analyses of musical performances, seismic records, and astrophysical time series, revealing a common underlying structure. This work establishes demodulation as a general route to 1/f fluctuations, providing a simple and scalable explanation for its ubiquity in both natural and engineered systems.
\end{abstract}

\def\oldabstract{}

\maketitle
\tableofcontents

\section{Introduction\label{sec:introduction}}

1/f fluctuation, or pink noise, is a ubiquitous phenomenon widely observed and is characterized by the low-frequency power law $S(\omega)\propto\omega^{-\beta}$
in the power spectral density (PSD), where $\omega$ is frequency and  $\beta=0.5\sim1.5$
\citep{Ward2007}. 
These 1/f fluctuations appear everywhere in nature, such as in semiconductors, thin metals,
bio-membranes, crystal oscillators,  heartbeat rates, magnetoencephalography (MEG), and electroencephalograms (EEG) (brain), etc.
\cite{Milloti2002}. 
Furthermore, artificial systems, such as music sound data, often exhibit typical 1/f fluctuations. 
This ubiquity strongly suggests the existence of a common underlying physical mechanism, yet its precise nature remains a subject of debate.

Since the first discovery of 1/f fluctuations in 1925 in the voltage fluctuations in the vacuum tube\cite{Johnson1925}, diverse theoretical explanations have been proposed, including self-organized criticality, fractional Brownian motion, percolation models, and aggregation-relaxation processes\cite{Milloti2002}. 
While these frameworks successfully describe certain systems, they lack universality: each model tends to be domain-specific and often requires fine-tuning to reproduce empirical scaling exponents. A single, simple, and scalable physical principle that can generate 1/f spectra across such varied contexts has yet to be established.

However, we believe that the universally observed phenomena cannot originate from complex physics but from a simple, ubiquitous physics.  
Therefore, we sought to investigate the fundamental origin of 1/f fluctuations, which reflects a simple physical principle. 

In our previous publications\cite{Morikawa2021}\cite{Morikawa2023}, we have proposed that the origin of all 1/f fluctuations is the wave beat, or amplitude modulation. In particular, if many waves with frequencies closer to each other can exhibit signals with arbitrarily low frequencies. 
Then, we attempted to verify and extend this proposal by applying it to earthquakes \cite{Nakamichi2023}, solar flares\cite{Morikawa2023-solar}, electric currents\cite{Morikawa2023-current}, and other phenomena. These systems had been pretty complicated for straightforwardly verifying this simple physics.  

In this work, we propose amplitude modulation (AM) combined with demodulation (DM) as a universal route to 1/f fluctuations.
In our framework, the 1/f scaling is not an intrinsic property of the raw waveform itself; rather, it emerges in the demodulated amplitude envelope, which is extracted through nonlinear transformations such as squaring, rectification, or thresholding. This viewpoint is motivated by empirical observations, particularly in music\cite{Voss1975}, where 1/f scaling is absent in the raw acoustic waveform but appears robustly after such transformations. We interpret this as evidence that the fluctuation resides in a 'hidden channel' (the envelope), which is only revealed by demodulation.

The key physical requirement for this AM/DM route is frequency accumulation, i.e., the presence of many oscillatory components with closely spaced frequencies, resulting in slow envelope modulations through wave beating. 
We identify two distinct yet complementary physical processes that naturally generate such frequency accumulation:
\begin{enumerate}
    \item{synchronization:}
    Stochastic synchronization among interacting oscillators leads to recurrent cycles of phase alignment and dispersion. We model this process using an extended stochastic Kuramoto framework capable of producing persistent low-frequency envelope variations without requiring fine-tuning to the classical synchronization threshold.
    \item{resonance:}
    Resonance-driven spectral shaping, in which environmental or structural eigenmodes selectively amplify certain frequencies, creating envelope modulations even in the absence of direct coupling among oscillators.
\end{enumerate}

Our approach offers three major departures from previous works:
\begin{itemize}
\item
    It unifies synchronization-based and resonance-based origins of 1/f noise within a single dynamical framework.
\item
    It shows that criticality is not a prerequisite: 1/f spectra arise over broad parameter ranges.
\item
    It demonstrates the cross-domain applicability of the AM/DM mechanism through numerical simulations and analyses of real-world data from orchestral and solo musical performances, seismic records, and astrophysical time series.
\end{itemize}

The remainder of this paper is organized as follows. 
Section II summarizes empirical observations and unresolved puzzles regarding 1/f fluctuations in musical signals. These turn out to be good hints for finding a simple physics for the origin of 1/f fluctuations. 
Section III introduces the stochastic Kuramoto model as a description of synchronization-induced AM and presents numerical results. 
Section IV develops a resonance-based AM model and validates it against measured acoustic environments. 
Section V integrates the two mechanisms into a unified framework, exploring their interplay in generating 1/f fluctuations. 
Section VI extends the analysis to natural and physical systems beyond music, illustrating the universality of the AM/DM route. 
Finally, Section VII discusses implications, limitations, and potential extensions of the present theory.


\section{Mysteries and some hints for music 1/f fluctuations\label{sec:mysteries}}

\begin{figure}
\includegraphics[width=9cm]{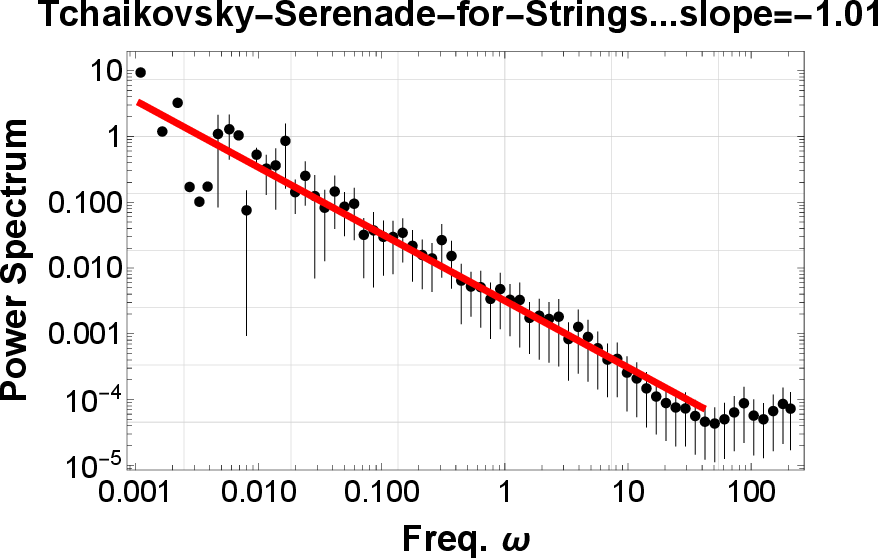}
\caption{
PSD of the sound data from Tchaikovsky-Serenade-for-Strings-Ozawa-Saito-Kinen orchestra \cite{Ozawa1992}.
The total data points are 80236463, which is divided into 802364 sections, and in each
of which, the sum of each data point squared corresponds to the sampling rate of 441 per second. 
The total duration is 1819.42 seconds. 
All the power spectral density data are binned with equal size on a logarithmic scale. 
The red line represents the power law in low-frequency regions fit obtained by least-squares fits.
The range is chosen from the low-frequency end toward the frequency from which the original sound data or any white noise tends to dominate: around \(40\)Hz in the present case. 
The remaining PSD graphs in this paper are obtained in the same method. 
The graph displays a power law with an index of $-1.01$ for more than four digits
The uniformity of the time series data may be of concern. 
Actually, in general, the data is not uniform, and the 1/f fluctuation property is not steady; thus, the PSD here describes the average property of the system. 
However, this non-uniformity itself is a powerful tool for observing the system much further. See section 7 Outlook and Fig.\ref{fig7}. 
}
\label{fig1}
\end{figure}

\begin{figure}
\includegraphics[width=9cm]{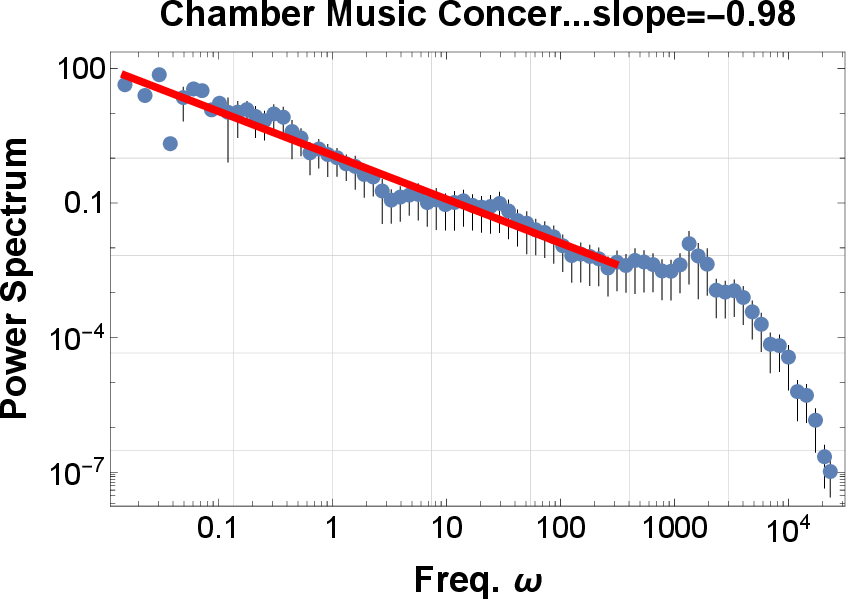}
\caption{
PSD of the sound data recorded at the Chamber Music Concert, OAC, 12 July 2023 by Kalliopi Petrou (soprano),
Stefano Menegus (Piano)\cite{Petrou2023}. 
The power-law fit (red) displays a power index of \(-0.98\) to more than five digits. }
\label{fig2}
\end{figure}

For a typical example of music sound, we show 1) the orchestra music concert\cite{Ozawa1992},
and 2) soprano and chamber music concert\cite{Petrou2023}. 
Their PSDs exhibit typical 1/f fluctuations, with indices of -1.01 in Fig.\ref{fig1} and -0.98 in Fig.\ref{fig2}, for more than 4 digits in both cases. 
Later, these examples turn out to be typical realizations of frequency accumulation: synchronization and resonance, respectively. 

We now examine 1/f fluctuations in music from very basic points.  
1/f fluctuation is enigmatic among various kinds of fluctuations. 
There are at least three mysteries in the 1/f fluctuations. 
\begin{enumerate}
    \item Low frequency signal continues \textbf{without limit toward low-frequency}: The power \(-1\) in PSD continues without bound. This fact indicates that the total power diverges since the energy per octave is equal. Furthermore, the system, if stationary, appears to possess an infinitely long memory, according to the Wiener-Khinchin theorem, which relates PSD to the time correlation function. 
    \item Arbitrary low-frequency signal arises from a \textbf{tiny system}.  In the ordinary argument, the system size determines the limiting frequency by the general order estimate. From the system size \(l\) and the typical wave speed \(v\), the maximum correlation time scale is of order \(l/v\). 
    However, the 1/f fluctuation in music violates this general rule.  The 100-meter music hall and the sound speed only yield a characteristic timescale of 0.3 seconds or several Hz.  On the contrary, the fluctuation of music 1/f remains for an hour or more. 
    Therefore, we speculate that 1/f fluctuation in music is not an intrinsic property but something secondary among many waves, such as interference between them.  
    \item  It is often the case that 1/f fluctuation appears in the PSD for the \textbf{data squared}; the original data never shows 1/f fluctuations.  Then, it would be unusual for the original signal to never show any remnant effect of the 1/f fluctuations. 
\end{enumerate}

Mystery 3 is a good starting point. We examine other operations on the original sound signal, such as arbitrary power operations, rectification, and thresholding, among others. 
Then, we found the following results of arbitrary power operation as in Table \ref{tab:oddeven}. 

\begin{table}
\caption{Measured spectral slopes under different n-th powers of data}
    \centering
    \begin{tabular}{ccccc}
    \toprule
         odd n-th power&  1&  3&  5& 7\\
         slopes&  0.84&  -0.20&  -0.15& 0.09\\
     \midrule   
         even n-th power&  2&  4&  6& 8\\
          slopes&  -1.27&  -1.15&  -0.89& -0.68\\
     \bottomrule     
    \end{tabular}
    \label{tab:oddeven}
\end{table}

These results naturally suggest that the square operation in item 3 is a kind of demodulation (DM) operation of the encoded 1/f fluctuations. Rapidly alternating positive and negative values in the original sound data cancel each other, leaving no information about the encoding.  

Suppose the two signals, with near frequencies (\(0<\lambda \ll \omega\)), are superposed, 
\begin{equation}
g=\sin((\lambda+\omega)t)+\sin((-\lambda+\omega)t)=\underset{A(t)}{\underbrace{2\cos(\lambda t)}}\sin(\omega t)
\label{eq:g}
\end{equation}
where the low-frequency signal \(A(t)\) is on the sine transportation wave: \(A(t) \sin (t \omega )\).
The n-th power of this becomes (\(n=1,2,...\))
\begin{equation}
\begin{split}
 g &= A(t) \sin (t \omega ) \\
 g^2 &= \frac{1}{2} A(t)^2\left(1- \cos (2 t \omega )\right) \\
 g^3 &= \frac{1}{4} A(t)^3\left(3  \sin (t \omega )- \sin (3 t
   \omega )\right) \\
 g^4 &= \frac{1}{8} A(t)^4 \left(3-4 \cos (2 t \omega)+  \cos (4
   t \omega )\right) \\
 g^5 &= \frac{1}{16} A(t)^5 \left(10\sin(\omega t)-5 \sin (3 t \omega)+  \sin (5
   t \omega )\right) \\
  & ...\\
\end{split}
\label{eq:powers}
\end{equation}
The isolated term of \(A(t)^n\) exists in the even-nth power, while \(A(t)^n\) mixes with the transportation wave in the odd-nth power.
The binomial expansion \(\sin(x)^n=((e^{ix}-e^{-ix})/(2i))^n\) generally proves this property. 
Furthermore, this even/odd-power property is also true for signals that include overtones in their timbre.  The similar expansion \((\sin(x)+\sin(2x))^n \) proves this fact.  
\textbf{Thus, we can naturally infer that the origin of 1/f fluctuation is the wave beat or the amplitude modulation (AM). We further hypothesize that 1/f fluctuation appears after any kind of demodulation operation (DM). }

We have checked other DM operations, such as absolute value and thresholding by the mean of the absolute value of the data, etc. Table\ref{tab:betavalues}.  All possible DM on the original sound data seem to extract 1/f fluctuations successfully.  

\begin{table}[ht]
\caption{Measured PSD spectral slopes under different signal transformations. 
PSD of the first 20 seconds sound data from Tchaikovsky-Serenade-for-Strings-Ozawa-Saito-Kinen orchestra \cite{Ozawa1992}. 
1/f fluctuation exhibits after all possible demodulation procedures, while it does not exhibit in the original sound data. 
\(H()\) is the Heaviside function. 
\(\mu\) is the mean of \(|\#|\). 
}
\begin{tabular}{lccc}
\toprule
Transformation & PSD indices \\
\midrule
Original Signal: \(\#\)      & 0.25  \\
Squared Signal: \(\#^2\)       & -1.25 \\
Absolute value: \(|\#|\) & -1.3 \\
Rectification: Max\((\#,0)\) & -1.28 \\
Negative-rectification: Min\((\#,0)\) & -1.27  \\
thresholding above \(\mu\): \# \(H(\#-\mu)\) &  -1.24 \\
anti-thresholding below mean but positive:\\
\# Max(\(H(\mu-\#),0)\) &  -0.89 \\
thresholding timing: \(H(\#-\mu)\) &  -1.1 \\
\bottomrule
\end{tabular}
\label{tab:betavalues}
\end{table}

If AM-DM were the origin of 1/f fluctuations, mystery two is naturally solved as a result of the \textbf{wave beat} associated with AM. Although the original data, as shown in Eq.(\ref{eq:g}), exhibits only two peaks in the PSD at the original frequencies, the square of the data exhibits an extra beat peak at an arbitrary low frequency, at \(2 \lambda \). 

The beat frequency  \(\lambda\) can be arbitrarily small. 
A typical example is the oldest electric instrument, the theremin, in which two nearly identical high-frequency generators exist; one is of fixed frequency, and the other is slightly controlled by adjusting the capacitance between the player's hand and the antenna.  
The wave beat between them yields an audible sound. 
This system also solves Mystery 1, as it yields arbitrarily low-frequency signals without bounds. 

However, a simple beat is not enough to yield the power law in PSD. 
We further need a systematic frequency accumulation for 1/f fluctuations.   

The most typical accumulation of frequencies would be the exponential form:  \(\omega=e^{-\kappa \xi}\), where \(\kappa\) is a constant and \(\omega\) is the frequency
\cite{Morikawa2023}. 
The positive variable \(\xi\) is uniformly random within some range with a constant probability \(p\).  
Then we immediately have the frequency distribution function inversely proportional to the frequency \(\omega\):
\begin{equation}
 P(\omega)=p\left|\frac{d \xi}{d \omega}\right|=\frac{p}{\kappa} \frac{1}{\omega},
 \label{Pomega}
\end{equation}
and the frequency difference (\(\Delta \omega\)) distribution function also becomes 
inversely proportional to \(1/\Delta \omega\). 
Based on this principle, we superpose many ($\mathit{N}$) sine waves with frequencies approaching each other exponentially,
\begin{equation}
\phi(t)=\sum_{i=1}^N\sin\left(2\pi\omega\left(1+ce^{-\xi_{i}}\right)t\right)
\label{eq:phi},
\end{equation}
where  $\xi_{i}$ is a uniform positive random variable and \(\omega,c\) are constants.
Numerical demonstration shows that the PSD of this wave superposition itself $\phi(t)$ does not 
display 1/f fluctuations, while  $\phi(t)^2$ it does. 
Similar frequency accumulation is expected also for the power-law approach
\cite{Morikawa2023}.

\section{Synchronization Mechanism: Orchestral Unison and the Stochastic Kuramoto Model\label{sec:synchronization}}

To model how 1/f fluctuations emerge from the collective behavior of multiple musical performers in unison, we employ a stochastic version of the Kuramoto model (SKM).
This model captures the essential dynamics of weakly coupled oscillators that tend to synchronize their phases despite individual differences in natural frequencies and noise perturbations.

\subsection{The Kuramoto Framework}

The classical Kuramoto model\cite{Kuramoto1975} describes the evolution of the phase \(\theta_i(t)\) of \(N\) coupled oscillators as:

\begin{equation}
\frac{d\theta_i(t)}{dt} = \omega_i + \frac{K}{N} \sum_{j=1}^{N} \sin(\theta_j(t) - \theta_i(t)),
\end{equation}
where \(\omega_i\) is the natural frequency of the \(i\)-th oscillator, and \(K\) is the global coupling strength. In a musical analogy, each \(\theta_i\) represents the phase of a performer's sound production, and the coupling term embodies auditory interaction or mutual adjustment among players.

The collective mode in the Kuramoto model is the mean of the phases:
\begin{equation}
r e^{i \psi}=\frac{1}{N} \sum_{i=1}^N e^{i \theta_i}. 
\label{eq:orderparameter}
\end{equation}
For the large \(N\) limit, the Kuramoto model has a critical point \(K_c\). 
The system does not synchronize at all (\(r=0\)) for \(K<K_c\), and gradually, \(r\) grows for \(K\) increases beyond \(K_c\)
\cite{Kuramoto1975}. 
Thus, synchronization or desynchronization is fixed by the parameters. In this stationary case, no large fluctuations nor 1/f fluctuations are observed.

To account for natural variation and timing instability among human performers, we introduce random noise\cite{Acebron2005}:
\begin{equation}\label{eq:SKM1}
\frac{d\theta_i(t)}{dt} = \omega_i + \frac{K}{N} \sum_{j=1}^{N} \sin(\theta_j(t) - \theta_i(t)) + \xi_i(t),
\end{equation}
where \(\xi_i(t)\) symbolically represents this randomness.  

However, this is not the ordinary Gaussian white noise. 
We introduce intermittent randomization events: at random times, each oscillator's phase \(\theta_i(t)\) is reassigned to a value drawn uniformly from \([0,1]\). This mimics an abrupt loss of synchrony followed by gradual re-synchronization via coupling. 
In an ensemble, each performer transitions from one note to the next. In the simplest model, this is captured by the stochastic phase reassignment, representing the onset of the next note without explicit rhythmic structure. The coupling term then realigns performers over time, producing cycles of partial synchronization and resynchronization.

Such resetting-type noise has been studied in non-equilibrium statistical physics and captures intermittent desynchronization phenomena more directly than continuous additive noise\cite{Evans2011}.
Sometimes, this resetting-type noise itself exhibits non-Gaussian properties. However, such an artificial effect is not present in our case, as the non-trivial PSD appears only after demodulation procedures. 

The above is not the sole extension of the Kuramoto model. Since we want to examine universal properties of synchronization and 1/f fluctuations, we also consider another form of extension, including the inertia term
\cite{Sonnenschein2013}
\cite{Rajwani2024}
: 
\begin{equation}\label{eq:SKM2}
\frac{d^2\theta_i(t)}{dt^2} =-\omega_i^2  \theta_i(t)+ \frac{K}{N} \sum_{j=1}^{N} \sin(\theta_j(t) - \theta_i(t)) + \xi_i(t).
\end{equation}

\subsection{From Phase to Sound Signal}

To construct a musical waveform from this model, we interpret each oscillator as emitting a sine wave whose instantaneous frequency is derived from its phase. Therefore, we use the imaginary part of  the above order parameter Eq.(\ref{eq:orderparameter}) as the indicator:
\begin{equation}\label{eq:x(t)}
x(t) = \frac{1}{N} \sum_{i=1}^{N} \sin(\theta_i(t)).
\end{equation}
The aggregated signal \(x(t)\) reflects the combined audio signal of the orchestra. Its envelope shows amplitude modulation arising from partial synchronization and beat phenomena among the oscillators.

\subsection{Numerical Simulation}
\begin{figure}
\includegraphics[width=9cm]{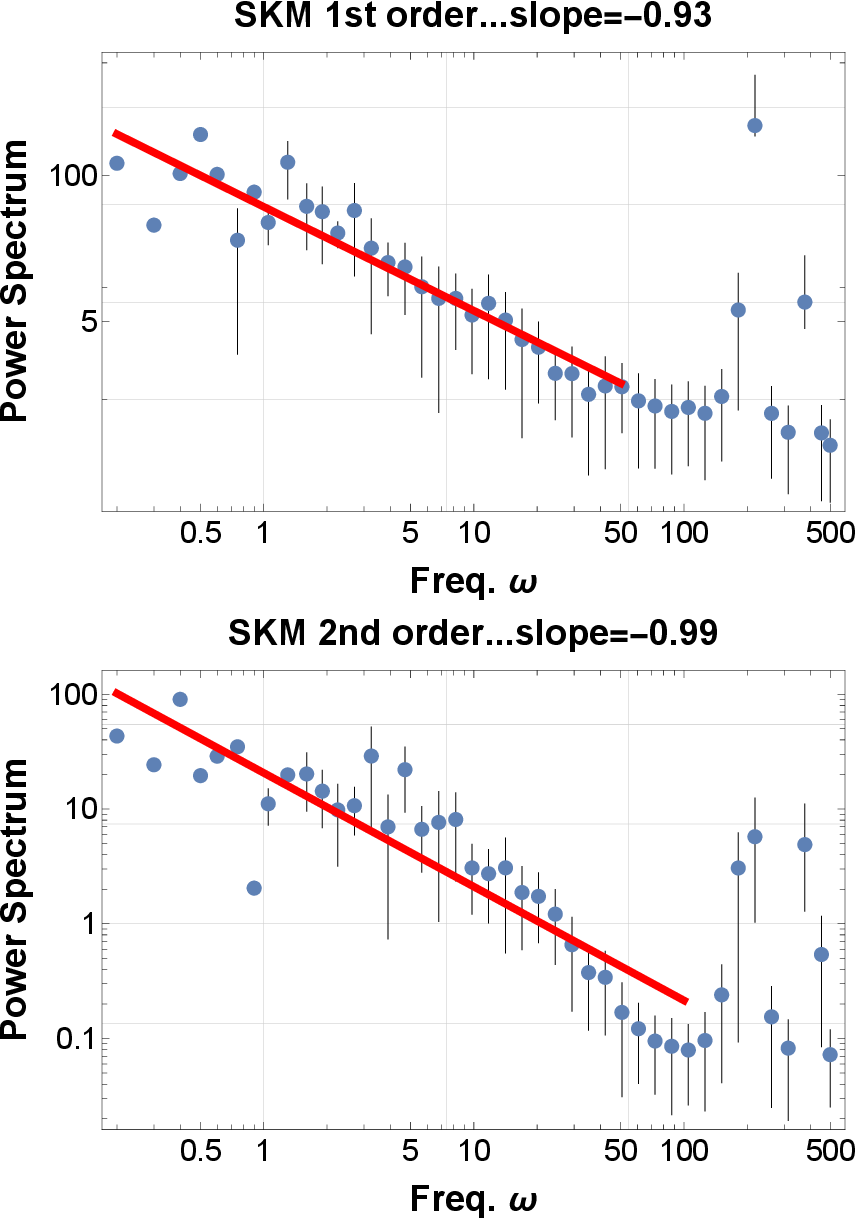}
\caption{\label{fig3}
\textbf{above)} PSD of time series \(x(t)\) calculated from the Kuramoto 1st order equation Eq.(\ref{eq:SKM1}).
Parameters are \(N=20, K=20\), the natural frequencies \(\omega_i\) are random uniformly distributed around \(100\) Hz within \(1\%\) range, and the time duration is \(10\). The random force effect, consistently in this paper, is realized by the random shift of the variables \(\theta_i\) of amount \([0,1]\) at a random timing with the interval \([0, 0.05]\).  
\textbf{below)} PSD of time series \(x(t)\) calculated from the Kuramoto 2nd order equation Eq.(\ref{eq:SKM2}).
Parameters are \(N=20, K=50\),
In both cases, 1/f fluctuations naturally arise without particular fine-tuning. The details are in Fig.\ref{fig6}.
\protect \\
}
\end{figure}

We first perform simulations of SKM for some values of  \(N, K\), where natural frequencies are uniformly distributed around 100 Hz within a 1\% range. After generating the waveform \(x(t)\), we compute the power spectral density (PSD) for both the raw and squared signals. There are several methods to simulate the random force effect, but the results on PSD do not change significantly. The details are in the captions of Fig.\ref{fig3}. 

While the raw signal exhibits only harmonic peaks due to clustering, the squared signal reveals a robust 1/f-type power law over several decades, as shown in Fig.~\ref{fig3}. 
Interestingly, this property is common to both models, Eqs. (\ref{eq:SKM1}) and (\ref{eq:SKM2}), displaying the generality of the mechanism of 1/f fluctuations from synchronization.  Further comparison is described in Section \ref{sec:Integration}. 

It is also important to point out the following additional properties of the simulation. 
\begin{enumerate}
    \item The partial sum over several \(\sin(\phi_i(t) )\) from Eq.(\ref{eq:x(t)}) still shows 1/f fluctuations, if squared.  
    \item Furthermore, a single variable \(\sin(\phi_i(t) )\) still shows 1/f fluctuations if self-superposed with the delayed data \(\sin(\phi_i(t-\tau) )\)and  squared.  This provides a notable contrast, as only a single variable \(\sin(\phi_i(t) )\) itself does not show 1/f fluctuations, even after being squared.  
    \item Even a bare superposition of the variables \(\theta_i(t)\) shows 1/f fluctuations if squared. 
\end{enumerate}
All of the above facts indicate that the 1/f fluctuation is quite robust, based on its interference with other variables and its history. 

\begin{figure}
\includegraphics[width=9cm]{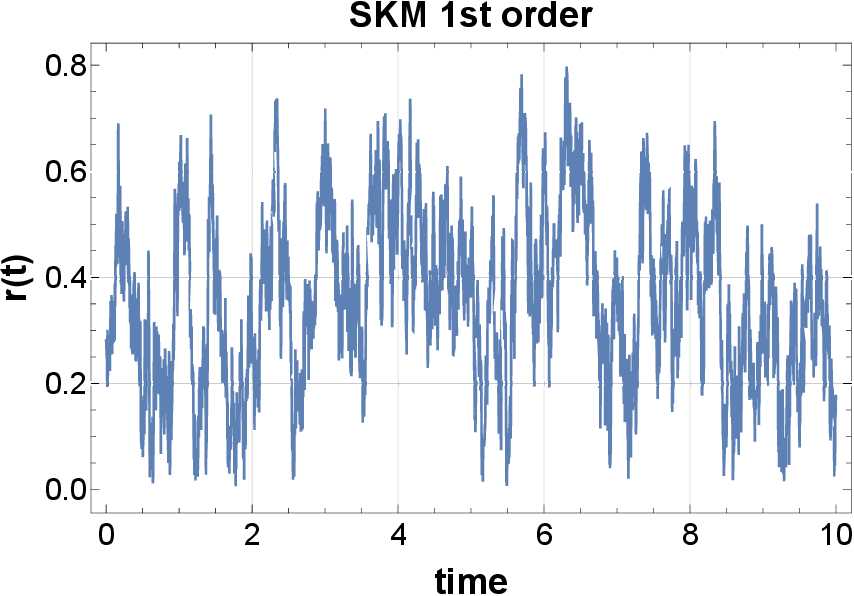}
\caption{\label{fig4}
Time evolution of the order parameter \(r(t)\) in Eq.\ref{eq:orderparameter} is strongly fluctuating. 
Parameters are the same as Fig.\ref{fig3} above. 
}
\end{figure}

We want to emphasize that the order parameter Eq.(\ref{eq:x(t)}) is strongly fluctuating as shown in Fig.\ref{fig4}, while \(N\) is finite; the low-frequency fluctuation is modulated in the order parameter. 
Furthermore, the system repeats synchronization and desynchronization as shown in Fig. \ref{fig4} in which the synchronization parameter \(r\) in Eq.(\ref{eq:orderparameter}) violently fluctuates.  

\subsection{Interpretation}

This simulation shows that when the input frequencies are narrow-band, the collective dynamics of perpetual synchronization and desynchronization generate low-frequency modulations. These modulations are then demodulated via a squaring operation, yielding 1/f fluctuations in the amplitude envelope.

Such a mechanism plausibly underlies the 1/f structure in orchestral recordings, where numerous performers with slightly different tempi attempt to maintain unison.
The orchestra retakes notes at each pitch transition of the musical melody. In this case, the orchestra synchronizes again at that timing in unison. In this way, music is a series of new synchronizations over and over again.

\section{Resonance Mechanism: Solo Performance and Acoustic Environments\label{sec:Resonance}}

In contrast to orchestral unison, where amplitude modulation arises from synchronization among performers, solo performances often exhibit amplitude modulation due to \textit{resonance} with the surrounding acoustic environment. This section explores how such resonance-induced amplitude shaping contributes to the emergence of 1/f fluctuations when combined with demodulation.

\subsection{Resonant Amplification in Physical Systems}

Acoustic resonance occurs when a sound wave's frequency matches the natural frequency of a cavity or structure, resulting in a significant amplification of amplitude. The frequency response of such systems often follows a Lorentzian profile:

\begin{equation}
H(f) = \frac{1}{(f - f_0)^2 + (\gamma/2)^2},
\end{equation}
where \( f_0 \) is the resonant frequency and \( \gamma \) characterizes the bandwidth or damping factor. In concert halls or vocal tracts, multiple such modes overlap to create a composite filtering effect.

\subsection{Solo Sound and Resonance Characteristics due to Room Reverberation}

A solo instrument or voice, when played in a hall, is effectively filtered by the 
resonance Characteristics due to room reverberation (RR).
The resulting waveform becomes amplitude-modulated due to resonant peaks in the 
RR.

We first calculate the eigenmodes of a typical concert hall, Gro{\ss}er Musikvereinssaal (Goldener Saal) in Vienna, Republic of Austria\cite{Musikverein2025}. 
This hall is almost rectangular: Depth \(L_1=48.8\) meters, Width \(L_1=19.2\) meters, and Height \(L_3=17.75\) meters, thus the following simple eigenvalues are associated with the Helmholtz acoustic equation,   
\begin{equation}\label{eq:f}
f=\frac{v_s}{2} \sqrt{(\frac{n_1}{L_1})^2+(\frac{n_2}{L_2})^2+(\frac{n_3}{L_3})^2}
\end{equation}
where \(v_s=343\) m/sec is the sound speed and \(n_1,n_2,n_3\) run \(1,2,...N\). 
Then superposing the sine wave of these frequencies, 
\begin{equation}\label{eq:phi}
\phi(t) = \sum_{n_1,n_2,n_3=0}^N \sin(2\pi f t),
\end{equation}
and further adding the reflection waves with time delay from each direction, we have the natural acoustic signal in this hall as 
\begin{equation}\label{eq:phia}
\phi_a(t) =\sum_{m=1}^3\sum_{k=1}^M (1+k)^{-\alpha} \phi(t-\frac{L_m}{
v_s}k),
\end{equation}
where \((1+n)^{-\alpha}\) with \(\alpha=0.1\) is assumed to be the power reduction rate for each reflection.  
These settings are a simple trial for our idea of amplitude modulation, although there are many elaborations and research in the field of room acoustics\cite{Aymerich2025}\cite{Marshall2014}.    

\subsection{Numerical Illustration}

\begin{figure}
\centering
\includegraphics[width=9cm]{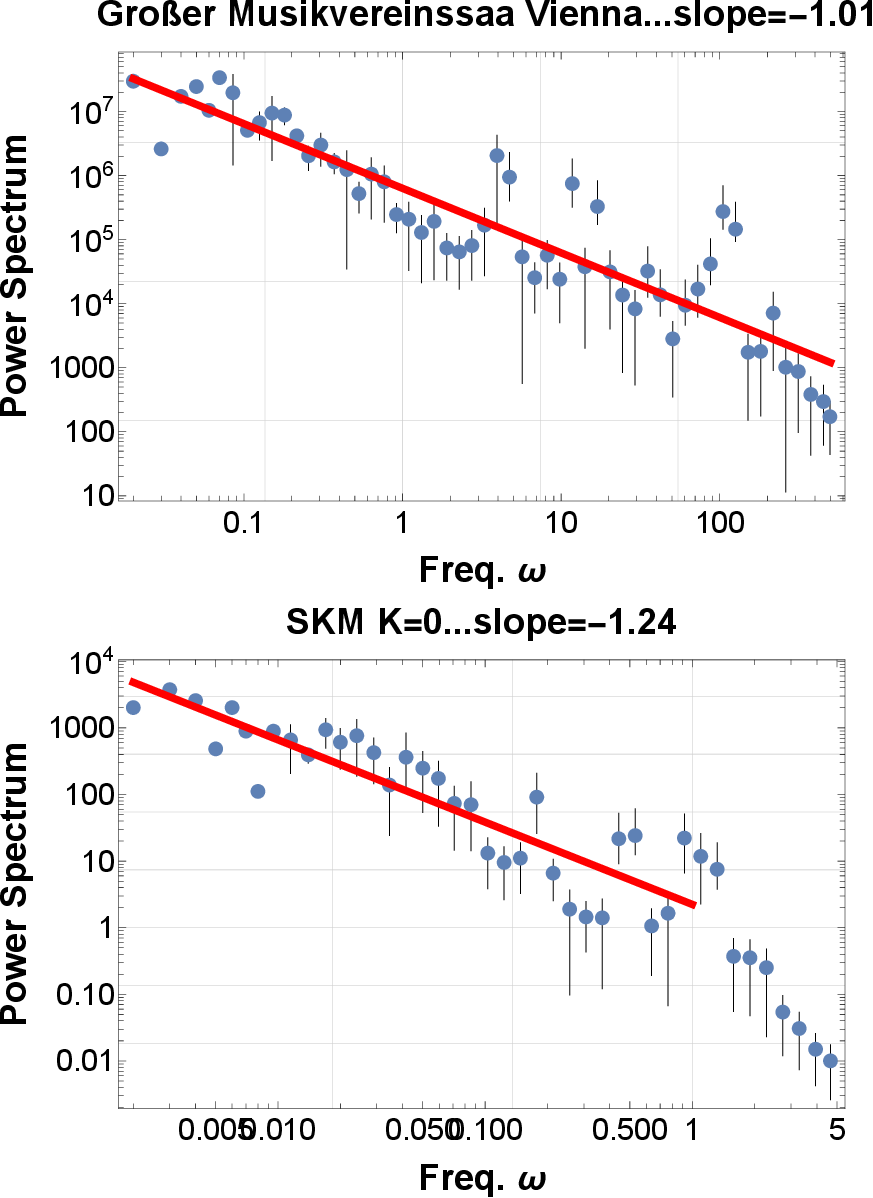}
\caption{
\textbf{above)} PSD of the Hall resonance. The typical case of Gro{\ss}er Musikvereinssaal (Goldener Saal) in Vienna, Republic of Austria. The resonating sound data is obtained by calculating the eigenfrequencies of the almost rectangular hall Eqs.\eqref{eq:f}, adding sine waves with the first 500 modes \eqref{eq:phi}, and further adding the acoustic reverberation of the hall \eqref{eq:phia}.  
\\
\textbf{below)} PSD of the time series created by SKM, disabling the interaction \(K=0\). Remarkably, 1/f fluctuation is obtained without synchronization. This fact is further analyzed in the next section. In this calculation, we considered a set of sine waves with adjacent frequencies in the range 1 Hz around the base frequency 100 Hz.  
}
\label{fig5}
\end{figure}

We simulate the resonating acoustic signal \(\phi_a(t)\) and its PSD, adopting  the \(500\) eigenfrequencies from the lowest, as well as \(M=20\) .  As before, PSD of the original signal \(\phi_a(t)\) reflects the eigenfrequencies and is always flat in low-frequency regions. 
On the other hand, the PSD of the absolute values of the signal \(\phi_a(t)\) shows 1/f fluctuations as in Fig.\ref{fig5} above. This demonstrates that the resonance of the eigenfrequencies exhibits 1/f fluctuations.

This resonance can also be expressed by SKM. 
Inputting many accumulating frequencies \(\omega_i\) and disabling the interaction \(K=0\), we obtain the frequencies agitated by the random resettings. In Figure \ref{fig5} below, we demonstrated PSD for the accumulating frequencies. As always, the raw waveform retains the flat spectral shape, while the absolute values of the data clearly exhibit a 1/f-like spectrum over a wide frequency range.

\subsection{Interpretation}

This supports the hypothesis that amplitude modulation from acoustic resonance plays a significant role in shaping the temporal structure of solo performances. As with synchronization-based AM, the 1/f structure only becomes evident after demodulation, reinforcing the central role of nonlinear envelope extraction in revealing 1/f fluctuations.

Such modulation mechanisms are ubiquitous in acoustics
found in caves, temples, cathedrals, and natural environments- contributing to the universal presence of 1/f fluctuations in sound. Some examples are in the Appendix. 

\section{Integration of Synchrony and Resonance: Not a Dichotomy\label{sec:Integration}}
\label{sec:5}
Thus far, we have presented synchronization and resonance as two distinct physical mechanisms for generating amplitude modulation (AM), which in turn leads to 1/f fluctuations upon demodulation. In this chapter, we emphasize that these mechanisms are not mutually exclusive; rather, they often coexist and interact synergistically in natural and musical systems.

\subsection{Unified View of Frequency Accumulation}

Both synchronization and resonance serve as processes of \textit{frequency accumulation}: 
\textbf{Synchronization} aggregates oscillators around a mean frequency through dynamic coupling.
\textbf{Resonance} enhances energy transfer at particular frequencies through structural filtering.

Each mechanism shapes the amplitude envelope of a waveform, leading to low-frequency AM components. These components manifest as 1/f fluctuation only after demodulation.
Even in orchestral unison, performers play in acoustic spaces that resonate with their sound. Likewise, a solo singer naturally synchronizes breathing and phrasing with accompanists or room acoustics.

\subsection{SKM description of frequency accumulation}

The spread spectrum of eigenfrequencies of the hall, instruments, and human body can be expressed by the set of frequencies \(\omega_i\), and the natural nonlinearity of the system necessarily introduces the finite interaction \(K\) in SKM Eq.\ref{eq:SKM1}.     

Further complications arise from the tone color or timbre that is intrinsically associated with the instruments or the singers. The sound color is determined by the unique combination of harmonics or overtones present in a sound wave, along with their relative intensities and how they change over time. Moreover, time delay and nonlinear frequency changes at the wall reflection, or the spatial extension of the instruments in the orchestra, may affect resonance and synchronization. 

Contrary to the above uncontrollable settings, musicians actively control the frequency vibrations (vibrato),  chest-voice/head-voice/humming resonance, glissando/legato, making the resonance and synchronization process quite complicated. 

We cannot perform a systematic analysis including the above effects in this paper. 
Instead, we aim to demonstrate how resonance and synchronization are incorporated into SKM. 
As we have already discussed, KM itself cannot describe the intermittent repetition of on-and-off synchronization, and cannot yield 1/f fluctuations.

\subsection{Phase-Amplitude Interaction Map}

We examined the extent to which both resonance and synchronization cooperatively yield 1/f fluctuations. 
The strength of synchronization is expressed in the coupling strength \(K\) in SKM, and that of resonance in the number of oscillators, N. 

\begin{figure}
\centering
\includegraphics[width=9cm]{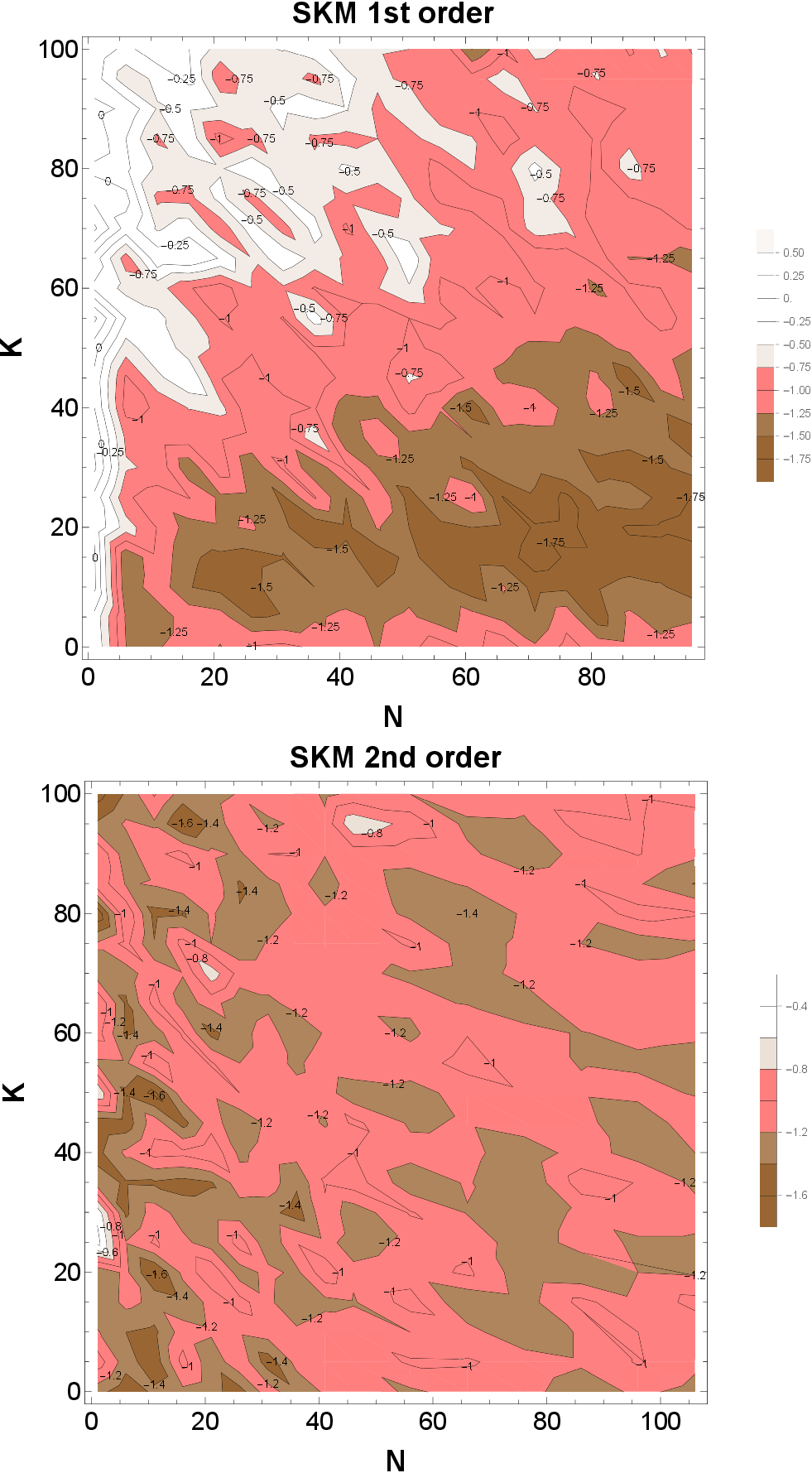}
\caption{
Density plots of PSD indices calculated from Stochastic Kuramoto models (SKM) of the 1st order (above), and the second order (below), by changing the parameters \(N:1\sim100\) and \(K:0\sim100\). 
1/f fluctuations regions (pink), marked by pink (indies \(\sim -1 \pm 0.2\)), are widely observed among Brown noise regions (brown) and White noise regions (white). 
\\
\textbf{above)} In this 1st order SKM case, the 1/f fluctuation is manifest around the diagonal (\(K\sim N\)).
These cases correspond to the cooperation of synchronization and resonance. Interestingly, 1/f fluctuations are also observed in another region, beyond the valley from those lines, along with the horizontal axes (\(K\sim 0\)). This case corresponds to the resonance-dominant region for 1/f fluctuations.
\\
\textbf{below)} The region of 1/f fluctuation is extended than the 1st order SKM. 
}
\label{fig6}
\end{figure}

In Fig.\ref{fig6} above, we plot the power PSD indices of the time series created by the 1st order SKM Eq. \ref{eq:SKM1}, varying the parameters \(K\) and \(N\).  As is evident, 1/f fluctuations, indices around \(-1\), naturally arise for \(N\approx K\), i.e., in the case that resonance and synchronization work in balance. 
On the other hand, stronger resonance \(N > K\) yields smaller power indices, resembling Brownian motion, while stronger synchronization \(N < K\) yields larger power indices, resembling white noise.  

In Fig.\ref{fig6} below, we plot the same for the 2nd order SKM Eq.\ref{eq:SKM2}. 
Comparing the PSD-indices distributions of the two figures, it appears that the region of 1/f represented by the second-order derivative is wider, but since the number of derivatives in the equations is different, a simple comparison would not be possible.

The above results clarify that the region of 1/f fluctuations is wide in both SKM models. 
The order parameter Eq.\ref{eq:orderparameter} wildly fluctuates all the time Fig.\ref{fig4}. 
Therefore, the distinction between synchronized and unsynchronized regions is meaningless. 
Thus, we propose the concept of \textbf{"dynamical synchronization"}, in which both synchronization and resynchronization alternate rapidly
\footnote{This is also exactly the case of economics, where the synchronization between many economic activities and the crash repeats almost forever. 
}  

This perspective reframes 1/f fluctuation not as the result of a single dominant mechanism but as an emergent signature of systems where multiple forms of convergence cooperate.

\subsection{Conclusion}

The dichotomy between synchrony and resonance is artificial. Many real-world systems exhibit features of both. 
Our framework acknowledges their interplay and offers a more comprehensive model for understanding the origin of 1/f fluctuations.


\section{Beyond Music: 1/f Fluctuations in Natural and Physical Systems\label{sec:beyond}}

While this study originates from the analysis of musical signals, the proposed AM/DM framework is not specific to music. In this chapter, we demonstrate that similar mechanisms underlie the 1/f fluctuations observed across various natural and physical systems, including geophysics, economics, and astrophysics.

\subsection{Seismic Fluctuations}

Earthquakes often exhibit 1/f fluctuation in their temporal power spectra\cite{Nakamichi2023}. The wave beat of multiple seismic modes (Earth Free Oscillation), each associated with the eigenmodes of the Earth, can be viewed as a form of amplitude modulation due to resonance. The superposition of these eigen modes, when squared or the occurrence timing series is extracted (which involves nonlinear demodulation), reveals 1/f spectra extending over several decades\cite{Nakamichi2023}.

\subsection{Solar and Space Phenomena}

Solar flares, solar wind turbulence, and magnetospheric activity are known to display 1/f spectra in their intensity profiles\cite{Morikawa2023-solar}. These systems naturally involve resonant modes (e.g., Alfv\'{e}n waves)\cite{Morikawa2023-solar} and synchronization phenomena (e.g., coupling of magnetic and electric current loops)\cite{Nakamichi2012}, making them fertile ground for the AM/DM perspective.

\subsection{Economic Time Series}

Stock prices, currency exchanges, and market indices often display long-range correlations and 1/f fluctuations. 
The economic system is always in a dynamic state of synchronization, where the synchronization between various circulations of matter and money and its repeated crashes.
Therefore, indices of economic activity may often exhibit 1/f fluctuations. 
Interestingly, economic 1/f fluctuation appears in PSD after the square operation on the detrended data, or volatility. 
This operation should correspond to demodulation, just as in music cases.
These fluctuations can be interpreted as emerging from the aggregation of trader behaviors (synchronization) and a reaction to exogenous news or events (resonance-like filtering)\cite{MorikawaY2025}.

\subsection{Astrophysical Signals}

Fast radio bursts (FRBs), soft gamma repeaters (SGRs), black hole-disk systems (AGNs), and neutron star oscillations, as well as solar flares\cite{Morikawa2023-solar} and seismic earthquakes\cite{Nakamichi2023}, show 1/f spectral trends in their emitted signals. These phenomena involve extreme versions of synchronization (e.g., macro-spin model)\cite{Nakamichi2012} and resonance (e.g., neutron star eigenmodes), as discussed in our related work under preparation\cite{Morikawa2025-NS}. 
In these cases, 1/f fluctuations are observed directly in their signal without a rectification procedure. This is because the demodulation processes are intrinsic to the systems, such as the magnetic reconnections and fault disruptions.  

On the other hand, the data on velocity fluctuations of the plasma wave around the Sun require a rectification process before exhibiting 1/f fluctuations\cite{Morikawa2023-solar}. 

\subsection{Conclusion}

Across these domains, the combination of a) multiple interacting sources with partially aligned frequencies (synchrony), and b) selective amplification due to system-specific resonance (resonance), produces amplitude modulations that, upon demodulation, yield 1/f fluctuations.

The presence of 1/f fluctuations across music, geophysics, finance, and astrophysics suggests a shared structural cause. Our AM/DM mechanism provides a unifying physics to interpret these phenomena, opening pathways for future interdisciplinary modeling and prediction.

\section{Conclusion and Outlook\label{sec:conclusion}}
In this work, we have proposed a universal mechanism for the emergence of 1/f fluctuations across diverse domains, rooted in two fundamental physical processes: amplitude modulation (AM) and demodulation (DM). Through theoretical derivation, numerical simulation, and cross-domain case studies, we demonstrated that 
a) AM arises from two main sources: synchronization among oscillators (e.g., orchestral unison) and resonance with environmental structures (e.g., solo performance in acoustic spaces).
b) DM, implemented through nonlinear transformations such as squaring, is essential for revealing latent 1/f spectral properties.
c) The AM/DM mechanism explains a wide range of 1/f phenomena in music, nature, and astrophysics.
d) Order parameters make it possible to capture the essence of the system dynamics by disregarding (ignoring) random fluctuations in individual components.
However, in reality, order parameters exhibit large fluctuations that behave as 1/f in the low-frequency range\cite{Yamaguchi2018}.

\subsection*{Outlook}

\begin{figure}
\centering
\includegraphics[width=9cm]{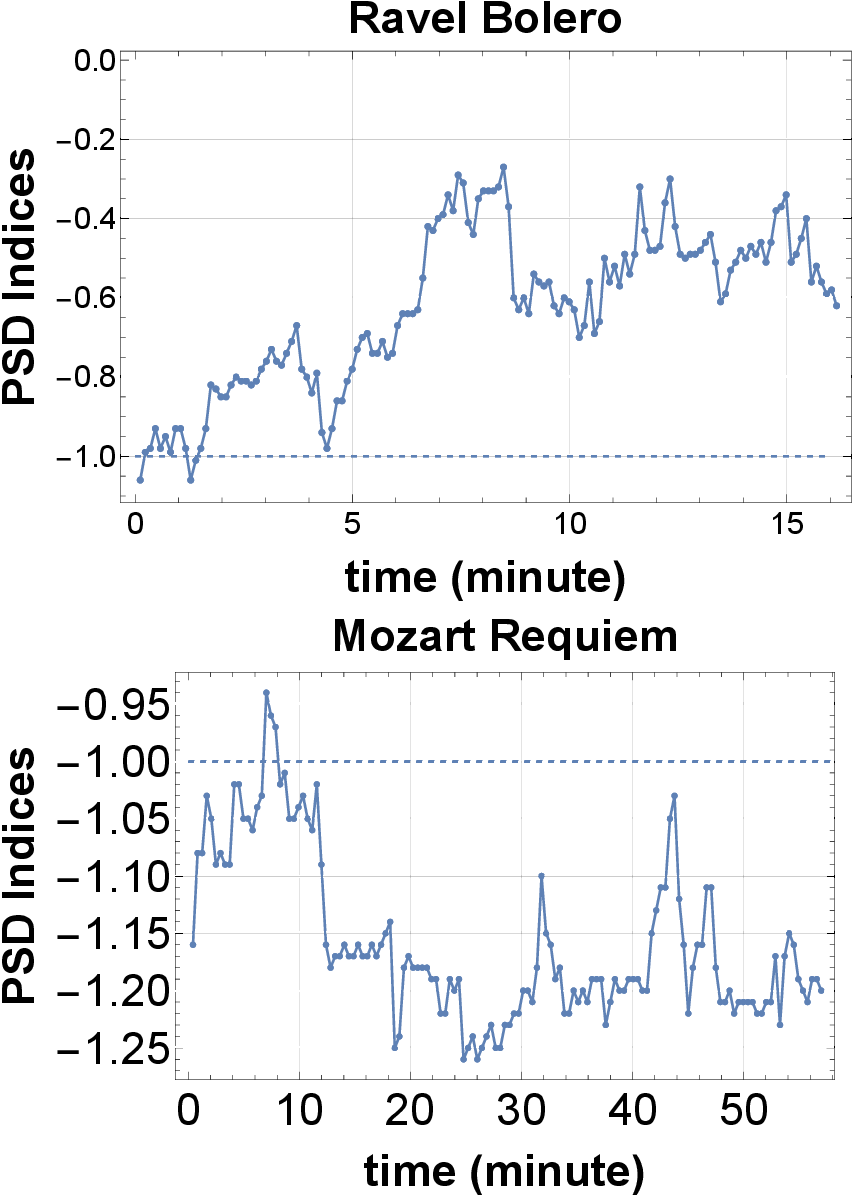}
\caption{
Examples of the time series of PSD indices around \(-1\). 
These analyses are useful for extracting the synchronization history of the system. \\
\textbf{above)} PSD indices time series for music, Ravel's Bolero
\cite{Ravel1928}, squared and analyzed in 2-minute segments, sliding every 6 seconds.
The history of PSD indices represents the degree of synchronization and does not reflect the sound volume.
\\   
\textbf{below)} PSD index time series for music, Mozart's Requiem\cite{Mozart1791}, squared and analyzed in seven-minute segments, sliding every 21 seconds. 
The first two sections (about ten minutes), composed by Mozart himself, exhibit more evident 1/f fluctuation (index \(\approx -1\)), while the rest of the sections, composed by the successor Sussmayr, deviate\cite{Wolff1994}. 
}
\label{fig7}
\end{figure}

The AM/DM framework opens several directions for future research:
\begin{enumerate}
    \item \textbf{Acoustic effects}:
    We have considered some simple features of music and sound. In reality, music is full of delicate sound effects that may affect the low-frequency fluctuations through resonance and synchronization. 
    They are the tone color or timbre,  time delay, spatial extension of sound field, vibrato, humming, glissando, legato,... 
    Furthermore, the recorded sound may be processed using various techniques, including reverb, chorus, delay, and distortion.
    We should integrate all of them for a complete resolution of musical 1/f fluctuation.  
    \item \textbf{Music pink noise from frequency modulation}:
    We have emphasized amplitude modulation (AM) in this paper and concentrated on the music performance. However, frequency modulation (FM) also yields a long-period structure from the individual short-period fluctuations, and music scores contain additional information, such as frequency time series and the rhythm of the music. 
    We want to explore 1/f fluctuations in music from a much wider perspective\cite{Voss1975}\cite{Voss1978}\cite{Levitin2012}\cite{Hsu1990}. 
    \item \textbf{PSD time series}: As the sound data is rich in data points, we can obtain local PSD indices by cutting the whole data into segments. These time series of PSD indices are useful for analysing the synchronization/resonance variation of the system. As shown in Fig.\ref{fig7}, we can detect a clear transition in musical mood (above), or a composer (below). 
    \item \textbf{Spatial 1/f fluctuation}: We can extend the ordinary notion of 1/f fluctuation in time domain to spatial domain as well: \(k^{-3}\) fluctuations for the wave number \(k\).
    A natural approach would be the Complex Ginzburg-Landau Equation (CGLE) \cite{Aranson2002}, the original equation of the Kuramoto model prior to phase reduction.  In this case, as in the time domain, spatial resonance and synchronization may characterize the long-distance correlations and \(k^{-3}\) fluctuations. 
    \\
    We must add another example of the frequency accumulation that we have not discussed in this paper. The primordial cosmic density fluctuations exhibit this type of spatial 1/f fluctuations, known as the Harrison-Zeldovich spectrum, and are derived from the infrared divergence\cite{Morikawa2022}\cite{Morikawa2023} in the same way as the electric current in QED\cite{Morikawa2023-current}. 
\end{enumerate}

\subsection*{Final Thoughts}

1/f noise, often treated as a mysterious or random phenomenon, emerges here as a natural consequence of structured temporal processes. Our model emphasizes that apparent complexity often arises from the interaction of simple, interpretable mechanisms: oscillation, modulation, and nonlinear transformation. Music may be the most intuitive manifestation of this principle, but it is far from the only one.

We hope this framework inspires further interdisciplinary exploration, where art and physics converge through the lens of temporal structure.


\section*{Appendix: Case Studies in Acoustic and Natural Environments}

In this appendix, we briefly introduce a series of case studies that highlight the versatility and explanatory power of the AM/DM framework. These real-world examples illustrate how 1/f fluctuations emerge in diverse acoustic settings and physical systems, extending beyond controlled laboratory or musical environments.

\subsection*{A. The Water Harp Cave at Hosen-in, Kyoto}
There is a variety of sound sources that yield 1/f fluctuations. 

\begin{figure}
\centering
\includegraphics[width=9cm]{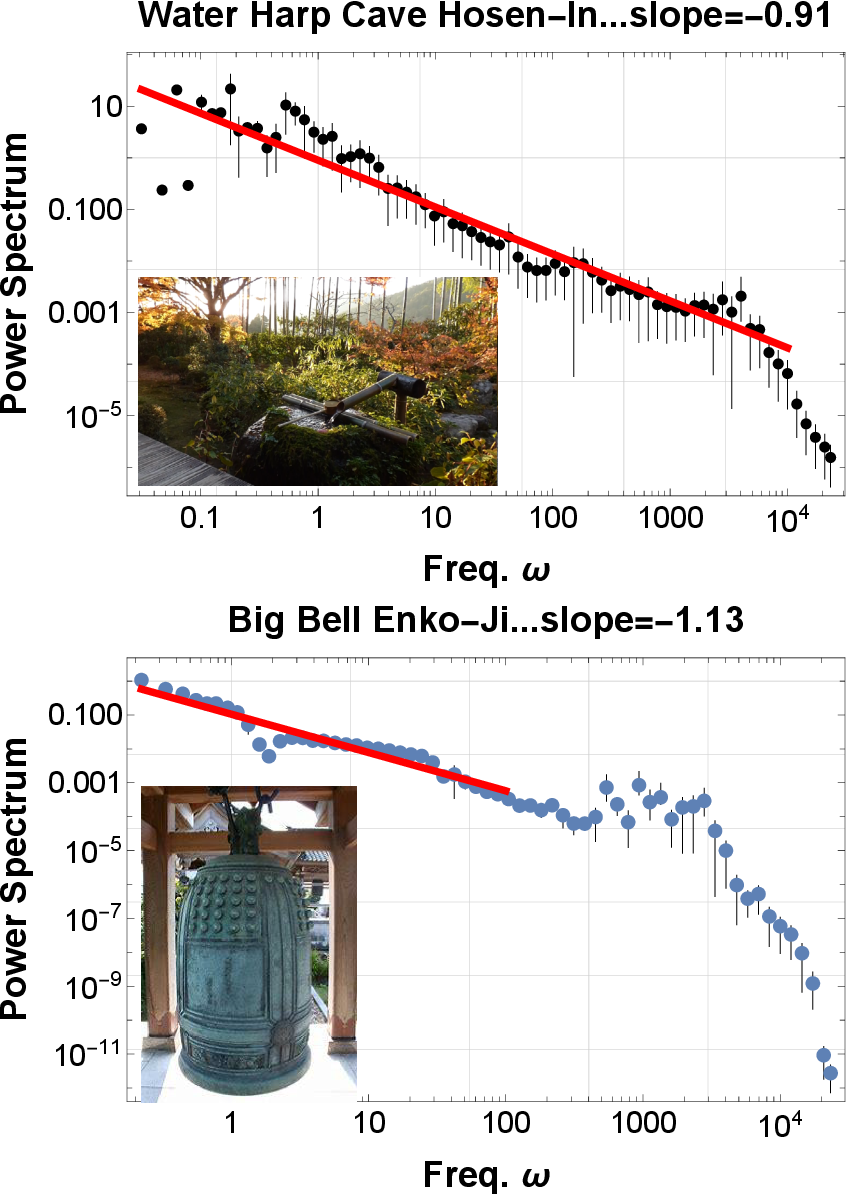}
\caption{
Kyoto sounds in temples. Sound resonance of devices in temples yields 1/f fluctuations. 
The data is squared before PSD analysis. 
\\
\textbf{above)} 
PSD of the sound data recorded on the Suikinkutsu, the water-harp cave at Hosen-In, Kyoto. 
The water from the basin is designed to drip into a 2-meter Mino-Yaki ceramic pot buried in the ground. 
The dripping water resonates and reverberates within the hollow of the pot, producing a clear melody like that of a koto.
Perpetual water drips activate the eigenfrequencies of the pot, like the random resetting triggers individual oscillators \(\theta_i\) in the Kuramoto model. 
\\
\textbf{below)}
PSD of the Big Bell sound data recorded at Enko-Ji temple, Kyoto. 
Initial strong strike activates the big-bell eigenmodes, which resonate and yield sound beats. 
This sound power emission resembles the energy emission by earthquakes; both are modulated by the wave beat of the system's eigenoscillations. 
}
\label{fig8}
\end{figure}

The Hosen-in temple in Kyoto features a "Water Harp Cave" that produces resonant sounds through underground water drips\cite{Hosen-In2025}. These perpetual water drops, falling into a hard ceramic vessel (Mino-yaki, approximately 2 meters in diameter) buried in the ground, excite resonant modes of the cavity. The result is an amplitude-modulated acoustic signal that, upon demodulation (e.g., via energy envelope extraction), reveals a 1/f spectral signature Fig.\ref{fig8} above.

Recordings from this site, when analyzed using squaring and spectral methods, show long-range correlation in the amplitude envelope, supporting our thesis that resonance and demodulation underlie 1/f fluctuation even in naturally occurring acoustic spaces.

Another soundscape is the Big Bell on the campus of Enkoji\cite{Enkoji2025} in the quiet northern Kyoto area.
PSD shows 1/f fluctuations(Fig.\ref{fig8} below). 

\subsection*{B. Cretan Sea Soundscape}

The recordings from the Cretan shore in Greece provide another compelling example. 
The periodic arrival and collapse of waves at the seashore at midnight yield a broadband acoustic signal.

\begin{figure}
\includegraphics[width=9cm]{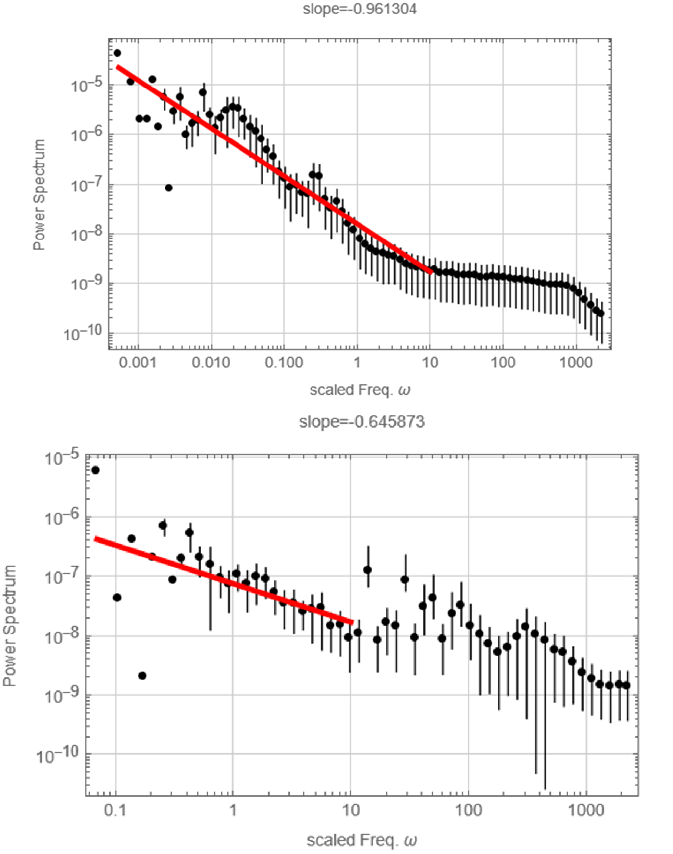}
\caption{
Cretan Sea sounds reflect the synchronization of many wave modes at the seashore.
\protect \\
\textbf{above)} 
PSD of the Cretan Sea sound data recorded at the Orthodox Academy of Crete
on July 13th, around 4am
before the down. 
\protect \\
PSD shows a power index of -0.96, indicating an almost 1/f fluctuation, in addition to some wave-specific quasi-periodicities.   
\protect \\
\textbf{below)} 
PSD of similar data on July 14th, around 6 am, 
after sunrise.  
The power index reduces to -0.65, although the frequency range is smaller. 
Sound contamination from the extremely loud cicada may disrupt the coherence and synchronization of the wave. 
The same may apply to the orchestral sound of Ravel's Bolero, in which a perpetual drum sound of indeterminate pitch may destroy the coherence and resonance of other instruments of determinate pitch. 
The PSD indices of Bolero's sound data do not exhibit 1/f fluctuations, unlike most other orchestral music.  
\protect \\
}
\label{fig9}
\end{figure}

Preliminary analysis indicates that after rectification or squaring, the amplitude fluctuations of this soundscape exhibit a 1/f spectrum on top of some periodic wave signals, Fig.\ref{fig9} above. 
This suggests that the marine acoustic wave naturally provides synchronization of many resonating waves that are favorable for the emergence of AM/DM-based 
1/f fluctuation.

However, after the dawn, this 1/f fluctuation is destroyed, resulting in a much flatter PSD, probably because the noisy cicadas contaminate the coherence of marine acoustics, Fig.\ref{fig9} below.  

Although temporally and physically distant from musical systems, these signals obey the same structural principles outlined in our AM/DM model.

\subsection*{C. Sound in Room and Hall}
We recorded a singing voice in a room and replayed it in both an ordinary office and a professional music hall at Kyoto-Sangyo University.  The PSD power index \(-1.48\) in the office (Fig.\ref{fig10} above), changed to  \(-1.06\) in the hall (Fig.\ref{fig10} below). 
This change may directly reflect the difference in the resonance effects in a room and a hall.  

\begin{figure}
\includegraphics[width=9cm]{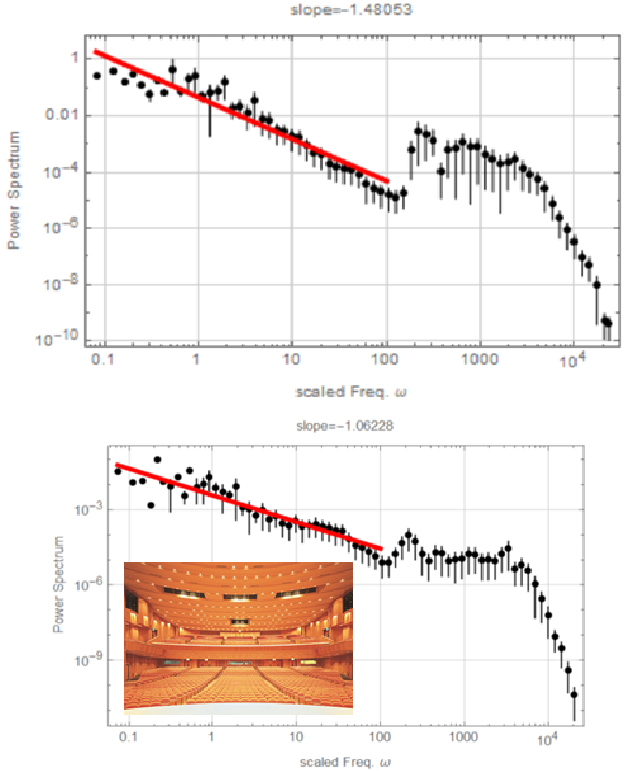}
\caption{
We first recoded the singing data, 'The Tulips Have Come into Bloom,' in an ordinary room. 
\protect \\
\textbf{above)} 
PSD obtained by replaying the recorded data in an office. 
The power index is \(-1.48\), not 1/f fluctuations. 
\protect \\
\textbf{below)} 
PSD obtained by replaying the recorded data at the stage in the Music Hall of Kyoto-Sangyo University. 
The PSD power index has changed to \(-1.07\), approaching the 1/f fluctuations. 
\\
We have performed the same experiments changing songs and styles, and found that PSD indices tend to approach -1 at this Hall, although the index were well below -1 in the office.     
The resonance of the hall may generally help to yield 1/f fluctuations. 
}
\label{fig10}
\end{figure}


\begin{acknowledgments}
AN would like to thank Manaya Matsui for helpful discussions and Junko Yamagata for beautifully singing the song 'Tulip'.
MM is grateful to Kalliopi Petrou for teaching him the basics of singing, including resonance techniques that utilize the cavities of the skull and chest.
They also thank the members of the remote lunch-time meeting for their many eye-opening comments and advice. 
In particular, many constructive criticisms by Akio Hosoya were useful in advancing our methods and extending our scope of music.  
\end{acknowledgments}

\bibliographystyle{unsrt}
\bibliography{pappinkmusic_references}
\end{document}